\documentclass[a4paper,11pt]{article}
\usepackage{epsfig}
\newcommand{\e}{\mathrm{e}}
\newcommand{\ep}{\epsilon}
\newcommand{\vev}[1]{\left\langle #1 \right\rangle}

\newcommand{\ffbox}[1]{\fbox{$\displaystyle #1$}}
\newcommand{\fmslash}[1]{\hbox{$#1$\kern-0.5em\raise0.3ex\hbox{/}}}

\newcommand{\Tr}{\mathrm{Tr}\,}
\begin{document}

\title{Renormalizable non-renormalizable theories \footnote{3 lectures
    given during September 2009 at the Feza G\"ursey Research
    Institute, Istanbul}} 

\author{H.~Sonoda\footnote{E-mail:
    \texttt{hsonoda@kobe-u.ac.jp}}\\ \textit{Physics Department, Kobe
    University, Kobe 657-8501, Japan}} 

\date{September 2009}

\maketitle

\begin{abstract}
  The general prescription for constructing the continuum limit of a
  field theory is introduced.  We then apply the prescription to
  construct the O(N) non-linear $\sigma$ model and the Gross-Neveu
  model in three dimensions using the large N expansions.  We compare
  these non-linear models with the linearized models, and show the
  equivalence. 
\end{abstract}

\tableofcontents

\newpage

\section{Lecture 1 -- Continuum Limits}

The purpose of the first lecture is to familiarize ourselves with the
concept of renormalization through concrete examples \footnote{The
  first lecture is almost identical to the first lecture given at the
  Pohang Winter School in February 2006.\cite{Sonoda:2006rr}}.  We try
to give a short summary of sect.~12 of \cite{Wilson:1973jj}. Before we
start, we should agree on the use of the Euclid metric as opposed to
the Minkowski metric.
\[
\eta_{\mu\nu} = \mathrm{diag} (1,-1,-1,-1) \longrightarrow
\delta_{\mu\nu} = \mathrm{diag} (1,1,1,1)
\]
Given an $n$-point Green function of a scalar field $\phi$
\[
\vev{\phi (\vec{x}_1, x_1^0) \cdots \phi (\vec{x}_n, x_n^0)}\quad
(x_1^0 > x_2^0 > \cdots > x_n^0)
\]
we obtain an $n$-point correlation function
\[
\vev{\phi (x_1) \cdots \phi (x_n)} \equiv \vev{\phi (\vec{x}_1, x_1^4)
  \cdots \phi (\vec{x}_n, x_n^4)}\quad
(x_1^4 > x_2^4 > \cdots > x_n^4)
\]
by the analytic continuation
\[
x_i^0 \longrightarrow - i x_i^4\qquad (i=1,\cdots,n)
\]
For example, the free scalar propagator
\[
\vev{\mathbf{T}\,\phi (x) \phi (0)} \equiv \int \frac{d^D p}{(2
  \pi)^D} \frac{i ~\e^{- i p x}}{p^2 - m^2 + i \ep} 
\qquad (p x \equiv p^0 x^0 - \vec{p} \cdot \vec{x})
\]
becomes
\[
\vev{\phi (x) \phi (0)} \equiv \int \frac{d^D p}{(2 \pi)^D}
\frac{\e^{i p x}}{p^2 + m^2}\qquad 
(p x \equiv p_4 x_4 + \vec{p} \cdot \vec{x})
\]
For large $|x| = r$, this damps exponentially
\[
\vev{\phi (x) \phi (0)} \sim \exp \left( - m r \right)
\]

The idea of a continuum limit is very simple and must be already
familiar to you.  We consider a theory with a momentum cutoff
$\Lambda$, meaning that the theory is defined only up to the scale
$\Lambda$.  For example, a lattice theory defined on a cubic lattice
of a lattice unit
\[
a = \frac{1}{\Lambda}
\]
has the momentum cutoff $\Lambda$.  (Figure 1)
\begin{figure}[t]
\begin{center}
\epsfig{file=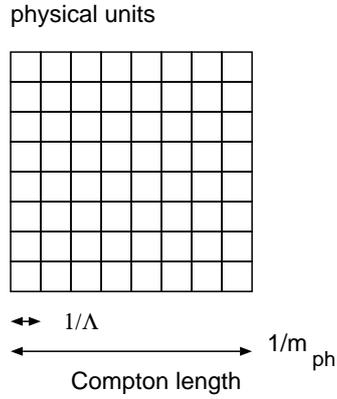}
\caption{A cubic lattice in physical units}
\end{center}
\end{figure}

\textbf{The continuum limit} is the limit
\[
\Lambda \to \infty
\]
and \textbf{renormalization} is a specific way of taking the continuum
limit so that the physical mass scale $m_{\mathrm{ph}}$, say the mass
of an elementary particle, remains finite.

\begin{figure}[t]
\begin{center}
\epsfig{file=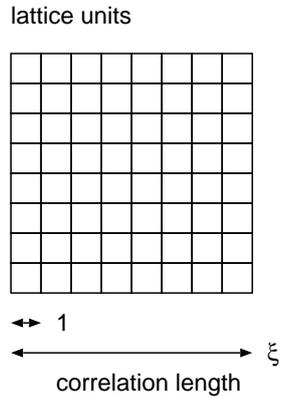}
\caption{A cubic lattice in lattice units\label{lattice}}
\end{center}
\end{figure}

From the viewpoint of a lattice theory, it is more natural to measure
distances in lattice units.  Hence, the lattice unit becomes simply
$1$. (Fig.~\ref{lattice})
In this convention, the Compton length or equivalently the inverse of
the physical mass is a dimensionless number $\xi$ called the
correlation length.  Therefore, we obtain
\[
\xi \, a = \frac{1}{m_{\mathrm{ph}}} \Longrightarrow
\ffbox{\xi = \frac{\Lambda}{m_{\mathrm{ph}}}}
\]
Clearly, $\xi \to \infty$ as we take $\Lambda \to \infty$ while
keeping $m_{\mathrm{ph}}$ finite.  Thus, as we take the continuum
limit, the lattice theory must obtain an infinite correlation length.

A lattice theory with an infinite correlation length is called a
\textbf{critical} theory.  Therefore, \textbf{to obtain a continuum
limit the corresponding lattice theory must be critical}.  Let us look
at examples.

\subsection{Ising model in $2$ dimensions}

The Ising model on a square lattice is defined by the action
\[
S = K \sum_{\vec{n} = (n_1,n_2)} \sum_{i=1}^2 \sigma_{\vec{n}}
  \sigma_{\vec{n} + \hat{i}}
\]
At each site $\vec{n}$ of the lattice, we introduce a classical spin
variable $\sigma_{\vec{n}} = \pm 1$.  The parameter $K$ is a
dimensionless positive constant, which we can regard as the inverse of
a reduced (i.e., dimensionless) temperature.
\[
K \sim \frac{1}{T}
\]
The partition function is defined by\footnote{We write $\e^S$ but not
  $\e^{-S}$ for the Boltzmann weight.}
\[
Z (K) = \sum_{\sigma_{\vec{n}} = \pm 1} \e^{S}
\]
and the correlation functions are defined by
\[
\vev{\sigma_{\vec{n}_1} \cdots \sigma_{\vec{n}_N}}_K =
\frac{\sum_{\sigma = \pm 1} \sigma_{\vec{n}_1} \cdots
\sigma_{\vec{n}_N} \e^{S}}{Z (K)}
\]

\begin{figure}
\begin{center}
\epsfig{file=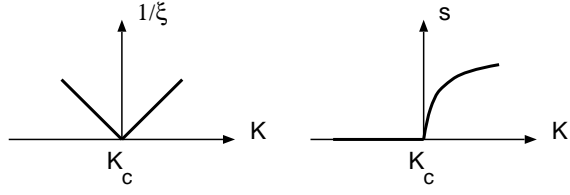}
\caption{The critical exponents $y_E, x_h$ characterize the
  correlation length $\xi$ and VEV $s$ near the critical point $K =
  K_c$.\label{critical2dising}}
\end{center}
\end{figure}

The action is invariant under the global $\mathbf{Z_2}$ transformation
\[
(\forall \vec{n})\quad \sigma_{\vec{n}} \longrightarrow -
\sigma_{\vec{n}}
\]  With respect to this symmetry, the model has two phases:
\begin{itemize}
\item High temperature phase $K < K_c$: the $\mathbf{Z}_2$ symmetry is
exact, and
\[
\vev{\sigma_{\vec{n}}} = 0
\]
\item Low temperature phase $K > K_c$: the $\mathbf{Z}_2$ symmetry is
  spontaneously broken, and
\[
\vev{\sigma_{\vec{n}}} = s (K) \ne 0
\]
\end{itemize}
For large $|\vec{n}|$, the two-point function behaves exponentially as
\[
\vev{\sigma_{\vec{n}} \sigma_{\vec{0}}}_K \sim \e^{-
  \frac{|\vec{n}|}{\xi}}
\]
This defines the correlation length $\xi (K)$.  At $K = K_c$ the
theory is critical with $\xi = \infty$.  Two \textbf{critical
exponents}
\[
\ffbox{y_E = 1} \quad \textrm{and} \quad
\ffbox{x_h = \frac{1}{8}}
\]
characterize the theory near criticality as follows
(Fig.~\ref{critical2dising}):
\begin{itemize}
\item As $K \to K_c$, the correlation length behaves as
\[
\xi \sim |K - K_c|^{- \frac{1}{y_E}} = \frac{1}{|K - K_c|}
\]
\item As $K \to K_c + 0$, the VEV behaves as
\[
s \sim |K-K_c|^{\frac{h}{y_E}} = |K-K_c|^{\frac{1}{8}}
\]
\end{itemize}
The correlation functions near the critical point $K \simeq K_c$ obeys
\textbf{the scaling law}:
\[
\ffbox{ \vev{\sigma_{\vec{n}_1} \cdots \sigma_{\vec{n}_N}}_K \simeq |K
  - K_c|^{N \frac{x_h}{y_E}} F_N^{\pm} \left( \frac{\vec{n}_1 -
    \vec{n}_N}{\xi}, \cdots, \frac{\vec{n}_{N-1} - \vec{n}_N}{\xi}
  \right)}
\]
where $\pm$ for $K > (<) K_c$.  For $N > 1$, the scaling law is valid
  only for large separation of lattice sites:
\[
|\vec{n}_i - \vec{n}_j| \gg 1 \quad (i \ne j)
\]

For $N=1$, the scaling law simply boils down to the power law behavior
of $s(K)$ near criticality.  For $N=2$, the scaling law gives
\[
\vev{\sigma_{\vec{n}} \sigma_{\vec{0}}}_K
 \simeq |K-K_c|^{\frac{1}{4}} F_2^\pm \left(\frac{\vec{n}}{\xi} \right)
\]
where $\xi \sim \frac{1}{|K-K_c|}$.  For the limit to exist as $K \to
K_c$, the function $F_2^\pm$ must behave like
\[
F_2^\pm (x) \sim x^{- 1/4}
\]
for $x \ll 1$.  Hence, at the critical point $K=K_c$, the two-point
function is given by the power law:
\[
\vev{\sigma_{\vec{n}} \sigma_{\vec{0}}}_{K_c} \sim
 \frac{1}{|\vec{n}|^{2 x_h}} = \frac{1}{|\vec{n}|^\frac{1}{4}}\quad
 (|\vec{n}| \gg 1)
\]
In fact this is another way of introducing the critical exponent
$x_h$.

The scaling law introduced above implies that we can renormalize the
Ising model to construct a scalar field theory as follows:
\[
\ffbox{
\vev{\phi (\vec{r}_1) \cdots \phi (\vec{r}_{N})}_{m; \mu}
 \equiv \lim_{t \to \infty} \e^{\frac{N t}{8}} \vev{\sigma_{\vec{n}_1
 = \mu \vec{r}_1 \e^t} \cdots \sigma_{\vec{n}_N = \mu \vec{r}_N
 \e^t}}_{K = K_c - \frac{m}{\mu} \e^{-t}}
}
\]
where both $m$ and $\mu$ have mass dimension $1$.\footnote{We chose a
  sign convention so that the $\mathbf{Z}_2$ is spontaneously broken
  for $m < 0$.}

Let us explain this formula in several steps:
\begin{enumerate}
\item $K = K_c - \frac{m}{\mu} \e^{-t}$ --- Hence, as $t \to \infty$,
  the theory approaches criticality.  The particular $t$ dependence
  was chosen so that $\xi \propto \e^t$.
\item Necessity of $\mu$ --- $\mu$ was introduced so that $\Lambda =
  \mu \e^t$.  Hence, the coordinate $\vec{r} = \vec{n} \frac{1}{\mu}
  \e^{-t}$ has mass dimension $-1$, and $m$ has mass dimension $1$.
\item Given an arbitrary coordinate $\vec{r}$, $\vec{n} = \mu \vec{r}
  \e^t$ is not necessarily a vector with integer components.  Since
  $\e^t \gg 1$, however, we can always find an integral vector
  $\vec{n}$ which approximates $\mu \vec{r} \e^t$ to the accuracy
  $\e^{-t} \ll 1$.
\item Applying the scaling law, we can compute the limit as
\begin{eqnarray*}
&&\vev{\phi (\vec{r}_1) \cdots \phi (\vec{r}_{N})}_{m; \mu} \\
&=&
\lim_{t \to \infty} \e^{\frac{N t}{8}} |K - K_c|^{\frac{N}{8}}
F_N^\pm \left( \frac{\mu \e^t |K-K_c|}{c_\pm} (\vec{r}_1 -
\vec{r}_N), \cdots \right)\\
&=& \left(\frac{m}{\mu}\right)^{\frac{N}{8}} F_N^\pm \left(
\frac{m}{c_\pm} (\vec{r}_1 - \vec{r}_N), \cdots \right)
\end{eqnarray*}
where we used
\[
\xi \simeq \frac{c_\pm}{|K-K_c|}
\]
Thus, the limit exists.  The limit depends not only on the mass
parameter $m$ but also on the arbitrary mass scale $\mu$.
\item \textbf{RG equation} --- The correlation function satisfies
\[
\vev{\phi (\e^{- \Delta t} \vec{r}_1) \cdots \phi (\e^{- \Delta t}
\vec{r}_{N})}_{m \e^t; \mu} = \e^{N \frac{\Delta t}{8}} \vev{\phi
(\vec{r}_1) \cdots \phi (\vec{r}_{N})}_{m; \mu}
\]
This implies that 
 the scale change of the coordinates
\[
(\forall i)\quad \vec{r}_i \longrightarrow \vec{r}_i \e^{- \Delta t}
\]
can be compensated by the change of the mass parameter $m$:
\[
m \longrightarrow m \e^{\Delta t}
\]
and renormalization of the field:
\[
\phi \longrightarrow \e^{\frac{\Delta t}{8}} \phi
\]
Hence, $y_E = 1$ is the \textbf{scale dimension} of $m$, and $x_h =
\frac{1}{8}$ is that of $\phi$. The general solution of the RG
equation is given by the scaling formula with $F_N^\pm$ as arbitrary
functions.
\item $\mu$ dependence (alternative RG equation) 
\[
\vev{\phi (\vec{r}_1) \cdots \phi (\vec{r}_{N})}_{m; \mu \e^{- \Delta t}}
= \e^{\frac{\Delta t}{8}} \vev{\phi
(\vec{r}_1) \cdots \phi (\vec{r}_{N})}_{m; \mu}
\]
This is obtained from the previous RG equation by dimensional
analysis.  The change of $\mu$ is compensated by renormalization of
$\phi$.  
\item Elimination of $\mu$ --- If we want, we can eliminate the
  arbitrary scale $\mu$ from the continuum limit by giving the mass
  dimension $\frac{1}{8}$ to the scalar field.  By writing
  $\mu^{\frac{1}{8}} \phi$ as the new scalar field $\phi$, we obtain
\[
\vev{\phi (\vec{r}_1) \cdots \phi (\vec{r}_{N})}_m
 = m^{\frac{N}{8}} F_N^\pm \left(
\frac{m}{c_\pm} (\vec{r}_1 - \vec{r}_N), \cdots \right)
\]
\end{enumerate}

Before ending, let us examine the short distance behavior using RG.
The RG equation can be rewritten as
\[
\vev{\phi (\e^{- t} \vec{r}_1) \cdots \phi (\e^{- t} \vec{r}_N)}_{m;
  \mu} = \e^{N \frac{t}{8}} \vev{ \phi (\vec{r}_1) \cdots \phi
  (\vec{r}_N)}_{m \e^{- t}; \mu}
\]
Hence, in the short distance limit, the correlation functions are
given by those at the critical point $m=0$:
\[
\vev{\phi (\e^{- t} \vec{r}_1) \cdots \phi (\e^{- t} \vec{r}_N)}_{m;
  \mu} \stackrel{t \gg 1}{\simeq} \e^{N \frac{t}{8}} \vev{\phi
  (\vec{r}_1) \cdots \phi (\vec{r}_N)}_{0; \mu}
\]
Especially for the two-point function, we obtain
\[
\vev{\phi (\e^{-t} \vec{r}) \phi (\vec{0})}_{m; \mu} \stackrel{t \gg
  1}{\simeq} 
\frac{\mathrm{const}}{(\mu r \e^{- t})^{\frac{1}{4}}}
\]

\subsection{Ising model in $3$ dimensions}

We can construct the continuum limit of the $3$ dimensional Ising
model the same way as for the $2$ dimensional one.  The only
difference is the value of the critical exponents $y_E$ and $x_h$.
\[
y_E \simeq 1.6,\quad \eta \equiv 2 x_h - 1 \simeq 0.04
\]
These are known only approximately.  $\eta$ gives the difference of
  the scale dimension of the scalar field from the free field value,
  and is called \textbf{the anomalous dimension}.

  In defining the continuum limit, we can give any engineering
  dimension to the scalar field.  Here, let us pick $\frac{1}{2}$, the
  same as for the free scalar field.  The two-point function can be
  defined as
\[
\vev{\phi  (\vec{r}) \phi (\vec{0})}_{g; \mu}
\equiv \mu \lim_{t \to \infty} \e^{(1 + \eta) t} \vev{\sigma_{\vec{n}
    = \mu \vec{r} \e^t} \sigma_{\vec{0}}}_{K = K_c -
  \frac{g}{\mu^2} \e^{- y_E t}}
\]
We have given the engineering $2$ to the parameter $g$ as for the
squared mass.

The two-point function obeys the following RG equation:
\[
\vev{\phi (\vec{r} \e^{- \Delta t}) \phi (\vec{0})}_{g \e^{y_E \Delta
 t}; \mu} = \e^{(1+\eta) \Delta t} \vev{\phi (\vec{r}) \phi
 (\vec{0})}_{g; \mu}
\]
This implies the short-distance behavior
\[
\vev{\phi (\vec{r} \e^{-t}) \phi (\vec{0})}_{g; \mu}
\stackrel{t \gg 1}{\simeq} \mathrm{const} \frac{\mu}{(\mu r
  \e^{-t})^{1 + \eta}}
\]

\newpage
\section{Lecture 2 --- O(N) non-linear $\sigma$ models in $D=3$}

On a cubic lattice, we define the O(N) non-linear sigma model by the
action
\[
S = K \sum_{\vec{n}=(n_1,n_2,n_3)} \sum_{i=1,2,3} \Phi^I_{\vec{n}}
\Phi^I_{\vec{n} + \hat{i}}
\]
where $\Phi^I_{\vec{n}}\,(I=1,\cdots,N)$ is an N-dimensional unit
vector defined on each lattice site, and the repeated $I$ is summed
over.  For $N=1$, we obtain the Ising model
\[
\Phi_{\vec{n}} = \pm 1
\]
The model has two phases:
\begin{enumerate}
\item $K < K_{cr}$ (high temperature phase) --- large fluctuations of
  fields are encouraged by small $K$, and the O(N) symmetry is exact.
\item $K > K_{cr}$ (low temperature phase) --- fluctuations of fields
  are suppressed, and $\Phi^I$ obtains a non-vanishing expectation
  value.  The O(N) symmetry is spontaneously broken down to O(N-1),
  resulting in N$-$1 massless Nambu-Goldstone boson fields.
\end{enumerate}
As the Ising model, the behavior of the non-linear $\sigma$ model near
criticality $K=K_{cr}$ is characterized by two critical exponents:
\begin{enumerate}
\item $y_E$ --- the correlation length is given by a constant multiple
  of $|K-K_{cr}|^{- \frac{1}{y_E}}$.
\item $x_h = \frac{1 + \eta}{2}$ --- for $K > K_{cr}$ near
  criticality, $\vev{\Phi^I} \propto (K-K_{cr})^{\frac{x_h}{y_E}}$.
\end{enumerate}

Following the general procedure explained in the previous section, the
continuum limit of the two-point function is obtained as
\[
\vev{\Phi^I (\vec{r}) \Phi^J (0)}_{g_E; \mu}
\equiv \mu \lim_{t \to \infty} \e^{2 x_h t} \vev{\Phi_{\vec{n} = \mu
    \vec{r} \e^t}^I \Phi^J_0}_{K = K_{cr} - \frac{g_E}{\mu^2} \e^{-
    y_E t}}
\]
where we have given the mass dimension $\frac{1}{2}$ to the scalar
field, and the mass dimension $2$ to the renormalized parameter $g_E$.
The limit depends on the arbitrary choice of the renormalization scale
$\mu$ as
\[
\left( - \mu \frac{\partial}{\partial \mu} + y_E g_E
  \frac{\partial}{\partial g_E} \right)  \vev{\Phi^I (\vec{r}) \Phi^J
  (0)}_{g_E; \mu} = \eta \vev{\Phi^I (\vec{r}) \Phi^J (0)}_{g_E; \mu}
\]

Alternatively, we can obtain the continuum limit using a theory
defined by a bare action with a momentum cutoff $\Lambda$.  Denoting
the physical length of a lattice unit as $a = \frac{1}{\Lambda}$, we
rewrite the lattice action as
\begin{eqnarray*}
S &=& - \frac{1}{2} K a^{-1} \, a^3 \sum_{\vec{n}} \sum_{I=1}^3 \frac{\left(
    \Phi^I_{\vec{n}} - \Phi^I_{\vec{n}+\hat{i}}\right)^2}{a^2}\\
&=& - \frac{1}{2 g_0} \int d^3 r\, \partial_\mu \Phi^I
(\vec{r}) \partial_\mu \Phi^I (\vec{r})
\end{eqnarray*}
where the dimensionless bare coupling and the field are defined by
\[
\frac{1}{g_0} \equiv K \frac{\Lambda}{\mu},\quad
\Phi^I (\vec{r}) \equiv \sqrt{\mu}\, \Phi^I_{\vec{n} = \Lambda \vec{r}}
\]
and the momentum cutoff $\Lambda$ is implied.  The fields satisfy the
non-linear constraint
\[
\Phi^I (\vec{r}) \Phi^I (\vec{r}) = \mu
\]
Using the above theory, the same continuum limit is obtained as
\[
\vev{\Phi^I (\vec{r}) \Phi^J (0)}_{g_E; \mu}
\equiv \lim_{\Lambda \to \infty} \left(\frac{\Lambda}{\mu}\right)^{2
  x_h} \vev{\Phi^I (\vec{r}) \Phi^J (0)}_{S (g_0)}
\]
where the bare coupling is chosen as
\begin{eqnarray*}
\frac{1}{g_0} &=& \left( K_{cr} - \frac{g_E}{\mu^2}
  \left(\frac{\mu}{\Lambda}\right)^{y_E} \right) \frac{\Lambda}{\mu}\\
&=&  \frac{1}{g_{0,cr}} - \frac{g_E}{\mu^2} \left(
  \frac{\mu}{\Lambda} \right)^{y_E - 1}
\end{eqnarray*}
where the critical value of the bare coupling given by
\[
\frac{1}{g_{0,cr}} = K_{cr} \frac{\Lambda}{\mu}
\]
is linearly divergent as $\Lambda \to \infty$.  To obtain a massive
theory ($g_E \ne 0$), we must tune $\frac{1}{g_0}$ to the accuracy of
$\left(\frac{\mu}{\Lambda}\right)^{y_E} \ll 1$.

\subsection{Perturbative non-renormalizability}

We would like to compute the two critical exponents $y_E, x_h$ using
the large N approximation.  Before doing that, let us study the theory
described by the bare action $S$ using perturbation theory for small
$g_0$.  Since we are in the low temperature phase, we take the
direction of the expectation value as the N-th, and express $\Phi^I$
in terms of $N-1$ independent fields $\phi^i\,(i=1,\cdots,N-1)$:
\[
\Phi^i = \sqrt{g_0} \, \phi^i,\quad
\Phi^N = \sqrt{\mu - g_0 \phi^i \phi^i}
\]
Substituting this into $S$, we obtain
\begin{eqnarray*}
S &=& - \frac{1}{2 g_0} \int d^3 x\, \left( g_0 \partial_\mu
  \phi^i \partial_\mu \phi^i + \frac{\left(g_0 \phi^i \partial_\mu \phi^i
    \right)^2}{\mu - g_0 \phi^i \phi^i} \right)\\
&=& - \frac{1}{2} \int d^3 x \, \left( \partial_\mu
  \phi^i \partial_\mu \phi^i + g_0 \frac{\left(\phi^i \partial_\mu \phi^i
    \right)^2}{\mu - g_0 \phi^i \phi^i} \right)
\end{eqnarray*}
The four-point interaction vertex is
\[
- \frac{1}{2} \frac{g_0}{\mu} \left( \phi^i \partial_\mu \phi^i
\right)^2
\]
where the coupling constant $\frac{g_0}{\mu}$ has mass dimension
$-1$.  The theory is obviously non-renormalizable.

Perturbative renormalization does not work because we are looking at
the neighborhood of the wrong point $g_0=0$.  We must look near the
critical point $g_0 = g_{0, cr}$ for renormalization.

\subsection{Large $N$ expansions}

To find $g_{0,cr}$ and the critical exponents $y_E, x_h$, we need an
approximation scheme.  We now take $g_0$ to be of order $\frac{1}{N}$,
\[
g_0 \propto \frac{1}{N}
\]
and expand the theory in powers of $\frac{1}{N} \ll 1$ \footnote{For
  the general method of 1/N expansions, see \cite{Coleman:1985as}.}
\footnote{See \cite{Arefeva:1979bd, Arefeva:1978fj} and references
  therein for extensive discussions of the use of 1/N expansions for
  renormalization of the 3d O(N) non-linear $\sigma$ model.}.  To make
the $N$ dependence manifest, we rewrite $g_0$ as $\frac{g_0}{N}$ from
now on.  So, the action is
\[
S = - \frac{N}{2 g_0} \int d^3 x\, \partial_\mu \Phi^I \partial_\mu
\Phi^I
\]

To introduce the large $N$ approximation, we first rewrite $S$ by
introducing an auxiliary field.  Using the integral representation of
the delta function
\[
\int_{-\infty}^\infty d\alpha\, \e^{i \alpha s} = 2 \pi \delta (s)
\]
we rewrite $S$ as
\[
S = - \frac{N}{2 g_0} \int d^3 x\, \partial_\mu \phi^I \partial_\mu
\phi^I + i \int d^3 x\, \alpha \left( \phi^I \phi^I - \mu \right)
\]
where $\phi^I$ are N independent fields satisfying no constraint, and
$\alpha$ is a real scalar field.  Replacing
\[
\phi^I \to \sqrt{\frac{g_0}{N}}\,\phi^I,\quad
\alpha \to \frac{N}{g_0} \alpha
\]
we obtain
\[
S = - \frac{1}{2} \int d^3 x\, \partial_\mu \phi^I \partial_\mu \phi^I
+ i \int d^3 x\, \alpha \left( \phi^I \phi^I - N \frac{\mu}{g_0}
\right)
\]
Let us define a functional of $\alpha (x)$ by
\[
\e^{f[\alpha]} \equiv \int [d\phi] \exp \left[ - \frac{1}{2} \int d^3
  x\, \partial_\mu \phi \partial_\mu \phi + i \int d^3 x\, \alpha
\left(\phi^2 - \frac{\mu}{g_0} \right) \right]
\]
so that the partition function is given as
\[
Z = \int [d\alpha] \prod_{I=1}^N [d\phi^I] \,\e^S = \int [d\alpha]\, 
\e^{N f[\alpha]}
\]
For $N \gg 1$, the functional integral over $\alpha$ is dominated by
the maximum of $f$, and the squared fluctuations of $\alpha$ are of
order $\frac{1}{N}$.

Let us suppose the maximum takes place at
\[
\alpha (x) = i m^2
\]
so that
\[
\frac{\delta f[\alpha]}{\delta \alpha}\Big|_{\alpha=im^2} = 0
\]
This gives
\[
\vev{\phi^2}_{m^2} = \frac{\mu}{g_0}
\]
where the VEV is evaluated using the free scalar theory of squared
mass $m^2$.  Hence, we obtain
\[
\int_p \frac{1}{p^2 + m^2} = \frac{\mu}{g_0}
\]
To evaluate the left-hand side, we must recall the momentum cutoff
$\Lambda$.  We then obtain
\[
\int_p \frac{1}{p^2 + m^2} = \frac{4\pi}{(2 \pi)^3} \int_0^\Lambda
dp\, \frac{p^2}{p^2 + m^2} = \frac{1}{2 \pi^2} \left( \Lambda - m
  \frac{\pi}{2} \right)
\]
where $m \equiv \sqrt{m^2} > 0$.  Thus, the minimum condition gives
\[
\frac{1}{2 \pi^2} \left( \Lambda - m \frac{\pi}{2} \right)
= \frac{\mu}{g_0}
\]
This gives
\[
\frac{1}{g_0} = \frac{1}{2 \pi^2} \frac{\Lambda}{\mu} - \frac{1}{4
  \pi} \frac{m}{\mu} > \frac{1}{g_{0,cr}}
\]
Hence, we obtain the critical coupling and the critical exponent $y_E$
as
\[
\frac{1}{g_{0,cr}} = \frac{1}{2 \pi^2} \frac{\Lambda}{\mu},\quad
y_E = 1
\]
We identify $g_E$ with $\frac{1}{4 \pi} m \mu$.  Since the scalar is
free in the large $N$ limit, we obtain
\[
x_h = \frac{1}{2}
\]

The critical value of $\frac{1}{g_0}$ is proportional to $\Lambda$ as
expected.  Since $m > 0$, we obtain 
\[
\frac{1}{g_0} < \frac{1}{g_{0,cr}}
\]
Hence, our results are valid only for the high temperature phase.

For the low temperature we must take account of the expectation value
\[
\vev{\phi^N} = \sqrt{\frac{N M}{8 \pi}}
\]
where $M$ is a positive mass scale parameterizing the VEV.  We also
expect the presence of N$-$1 massless Nambu-Goldstone bosons.  With
the shift
\[
\phi^N \to \sqrt{\frac{N M}{4 \pi}} + \phi^N
\]
we obtain
\[
S = - \frac{1}{2} \int d^3 x\, \partial_\mu \phi^I \partial_\mu \phi^I
+ i \int d^3 x\, \alpha \left( \phi^I \phi^I + 2 \sqrt{\frac{NM}{4
      \pi}}\, \phi^N + N \left(\frac{M}{4 \pi} - \frac{\mu}{g_0} \right)
\right)
\]
To get massless scalars, the saddle point must take place at
$\alpha=0$, and we obtain the condition
\[
\int_p \frac{1}{p^2} = \frac{\mu}{g_0} - \frac{M}{4 \pi}
\]
This gives
\[
\frac{\Lambda}{2 \pi^2} = \frac{\mu}{g_0} - \frac{M}{4 \pi}
\]
Hence,
\[
\frac{1}{g_0} = \frac{1}{2 \pi^2} \frac{\Lambda}{\mu} + \frac{1}{4
  \pi} \frac{M}{\mu}
\]
We find the same critical value $g_{0,cr}$ and exponent $y_E$.  Now,
we identify $g_E$ with $- \frac{1}{4 \pi} M \mu < 0$.  The scalar
fields $\phi^i\,(i=1,\cdots,N-1)$ are free \& massless in the large
N limit, and we obtain $x_h = \frac{1}{2}$.

\subsection{$\frac{1}{N}$ corrections}

The $\frac{1}{N}$ corrections are obtained from the fluctuations of
$\alpha$.  In the symmetric phase, we write
\[
\alpha = i m + \frac{\delta \alpha}{2 \sqrt{N}}
\]
Substituting this into the action, we obtain
\[
S = - \frac{1}{2} \int d^3 x\, \left(\partial_\mu \phi^I \partial_\mu
  \phi^I + m^2 \phi^I \phi^I \right) + \frac{i}{\sqrt{N}}
\int d^3 x\, \delta \alpha \left( \frac{1}{2} \phi^I \phi^I - N
  \frac{\mu}{2 g_0} \right)
\]
\begin{figure}[t]
\begin{center}
\epsfig{file=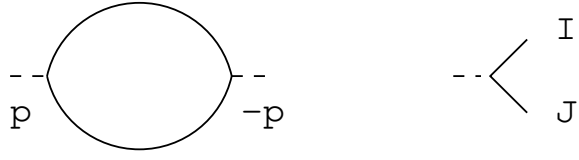, height=2cm}
\end{center}
\caption{2-point vertex for $\alpha$, and an interaction
  vertex\label{alpha2pt}}
\end{figure}
The two-point vertex of $\delta \alpha$ is obtained as (Fig.~\ref{alpha2pt})
\[
- \frac{1}{2} \int_q \frac{1}{q^2+m^2} \frac{1}{(p+q)^2 + m^2}
= - \frac{1}{8 \pi} \frac{1}{\sqrt{p^2}} \arctan \sqrt{\frac{p^2}{4 m^2}}
\]
Thus, the propagator of the $\alpha$ field is given by the inverse of
the above:
\[
\frac{8 \pi \sqrt{p^2}}{\arctan \sqrt{\frac{p^2}{4 m^2}}}
\]
The interaction vertex
\[
 i \frac{1}{\sqrt{N}} \delta^{IJ}
\]
is suppressed for large $N$ (Fig.~\ref{alpha2pt}), and the four-point
interaction vertex of the scalar field is given by
\[
- \frac{8 \pi}{N} \frac{1}{\arctan \sqrt{\frac{p^2}{4 m^2}}}
\]

In the broken phase, the analysis is a little more complicated since
we must treat $\phi^i$ and $\phi^N$ differently.  Rewriting 
\[
\alpha \to \frac{1}{2 \sqrt{N}}\,\alpha
\]
we obtain
\[
S = - \frac{1}{2} \int d^3 x \, \partial_\mu \phi^I \partial_\mu
\phi^I + \frac{i}{\sqrt{N}} \int d^3 x\, \alpha \left(
\frac{1}{2} \phi^I \phi^I + \sqrt{\frac{NM}{4 \pi}}\, \phi^N +
\frac{N}{2} \left(\frac{M}{4\pi} - \frac{\mu}{g_0}\right) \right)
\]
We find that $\alpha$ and $\phi^N$ mix.  By inverting two-point
vertices, we obtain the following propagators to leading order in 1/N:
\[
\left\lbrace
\begin{array}{c@{~=~}l}
\vev{\phi^N (p) \phi^N (-p)} & \frac{1}{\sqrt{p^2}} \cdot
\frac{1}{\sqrt{p^2} + \frac{4 M}{\pi}}\\
\vev{\phi^N (p) \alpha (-p)} & - i \frac{16 \sqrt{\frac{M}{4
      \pi}\,p^2}}{p^2 + \frac{4M}{\pi} \sqrt{p^2}}\\
\vev{\alpha (p) \alpha (-p)} & \frac{16 p^2 \sqrt{p^2}}{p^2 +
  \frac{4M}{\pi} \sqrt{p^2}}
\end{array}\right.
\]
Neither $\alpha$ nor $\phi^N$ correspond to stable particles.

\newpage
\section{Lecture 3 --- Gross-Neveu model in $D=3$}

The bare action of the Gross-Neveu model in 3 dimensions is given by
\footnote{The non-perturbative renormalizability of this model was first
discussed in \cite{Rosenstein:1988pt}.}
\[
S = - \int d^3 x \, \left[ \bar{\psi}^I \frac{1}{i} \sigma \cdot \partial
  \psi^I + \frac{g_0}{2 N} \left( \bar{\psi}^I \psi^I \right)^2\right]
\]
where we take
\[
g_0 > 0
\]
The action is invariant under the following $\mathbf{Z_2}$
transformation:
\[
\left\lbrace
\begin{array}{c@{~\longrightarrow~}l}
\psi^I (x) & \psi^I (-x)\\
\bar{\psi}^I (x) & - \bar{\psi}^I (-x)
\end{array}\right.
\]
Under this transformation, we find that $\bar{\psi}\psi$ changes sign:
\[
\bar{\psi}^I \psi^I (x) \longrightarrow - \bar{\psi}^I \psi^I (-x)
\]

The bare coupling $g_0$ has mass dimension $-1$, and the theory is
non-renormalizable perturbatively around $g_0 = 0$.  But the theory is
renormalizable if it has a non-trivial critical point.  For large
$g_0$, the interactions are suppressed, and for small $g_0$, the
interactions are enhanced.  Hence, we expect the existence of a
critical point $g_{0,cr}$ so that
\begin{enumerate}
\item $g_0 > g_{0,cr}$ --- $\mathbf{Z_2}$ is spontaneously broken, and
\[
\vev{\bar{\psi}^I \psi^I} \ne 0
\]
\item $g_0 < g_{0,cr}$ --- $\mathbf{Z_2}$ is exact, and 
\[
\vev{\bar{\psi}^I \psi^I} = 0
\]
\end{enumerate}

Given the critical exponents $y_E$, $x_h$, the continuum limit
is obtained as
\[
\vev{\psi^I (\vec{r}) \bar{\psi}^J (0)}_{g_E; \mu}
= \lim_{\Lambda \to \infty} \left(\frac{\Lambda}{\mu}\right)^{2
  (x_h-1)} \vev{\psi^I (\vec{r}) \bar{\psi}^J (0)}_{S}
\]
where we choose the bare coupling as
\[
\frac{1}{g_0} = \frac{1}{g_{0,cr}} + g_E
\left(\frac{\Lambda}{\mu}\right)^{1-y_E}
\]
where we give mass dimension 1 to $g_E$.  We wish to compute $y_E,
x_h$ using the large N approximation.

As a preparation, we introduce an auxiliary field $\alpha$ to rewrite
the action as follows:
\[
S = - \int d^3 x\, \left[ \bar{\psi}^I \frac{1}{i} \sigma
    \cdot \partial \psi^I + \frac{g_0}{2 N} \left( \bar{\psi}^I \psi^I \right)^2
      + \frac{1}{2} \left( \alpha + i \sqrt{\frac{g_0}{N}} \bar{\psi}^I \psi^I
      \right)^2 \right]
\]
Integrating out $\alpha$, we restore the original action.  Expanding
the gaussian term, we obtain
\[
S = - \int d^3 x\, \left[ \bar{\psi}^I \frac{1}{i} \sigma
    \cdot \partial \psi^I + \frac{1}{2} \alpha^2 + 
 i \alpha \sqrt{\frac{g_0}{N}} \bar{\psi}^I \psi^I \right]
\] 
Rescaling
\[
\alpha \longrightarrow \sqrt{N}\,\alpha
\]
we obtain
\[
S = - \int d^3 x\, \left[ \bar{\psi}^I \frac{1}{i} \sigma
    \cdot \partial \psi^I + \frac{N}{2} \alpha^2 + 
 i \alpha \sqrt{g_0} \,\bar{\psi}^I \psi^I \right]
\]
Hence, the vacuum functional integral is given as
\[
Z = \int [d\alpha] \prod_{I=1}^N [d\psi^I d\bar{\psi}^I]\, \e^S
= \int [d\alpha] \,\e^{N f[\alpha]}
\]
where
\[
\e^{f[\alpha]} \equiv \exp \left[ - \int d^3 x\, \frac{\alpha^2}{2}
\right] \int [d\psi d\bar{\psi}]\,
\exp \left[ - \int d^3 x\, \left( \bar{\psi} \frac{1}{i} \sigma
    \cdot \partial \psi + i \alpha \sqrt{g_0}\, \bar{\psi}
    \psi \right)\right]
\]
In the large N limit, we solve
\[
\frac{\delta f[\alpha]}{\delta \alpha} = 0 \Longrightarrow
\alpha = - i \sqrt{g_0} \vev{\bar{\psi} \psi}
\]

\subsection{Unbroken phase $g_0 < g_{0,cr}$}

\begin{figure}[t]
\begin{center}
\epsfig{file=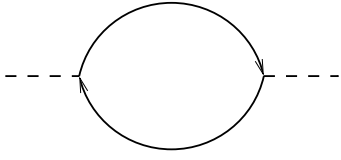, height=2cm}
\caption{Two-point interaction vertex for $\alpha$\label{GN-alpha2pt}}
\end{center}
\end{figure}

\begin{figure}[t]
\begin{center}
\epsfig{file=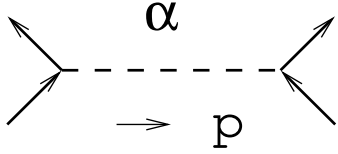, height=2cm}
\caption{The four-point interaction to leading order in
  1/N\label{GN-interaction}}
\end{center}
\end{figure}

In the unbroken phase $\vev{\bar{\psi}^I \psi^I}$ vanishes, and the
maximum of $f[\alpha]$ occurs at $\alpha = 0$.  The $\mathbf{Z_2}$
symmetry implies that the fermions are massless.  The two-point
interaction vertex of $\alpha$ is given by (Fig.~\ref{GN-alpha2pt})
\[
(-) (-i)^2 g_0 N \int_q \Tr \frac{1}{\sigma \cdot q}
\frac{1}{\sigma \cdot (p+q)}
= N g_0 \left( \frac{\Lambda}{\pi^2} - \frac{\sqrt{p^2}}{8}
\right)
\]
Thus, with the gaussian term, the entire two-point vertex becomes
\[
N \left( 1 - g_0 \left( \frac{\Lambda}{\pi^2} - \frac{\sqrt{p^2}}{8}
\right) \right)
= N g_0 \left( \frac{1}{g_0} - \frac{\Lambda}{\pi^2} +
  \frac{\sqrt{p^2}}{8} \right)
\]
Hence, with a positive mass parameter $M$, we must find
\[
\frac{1}{g_0} - \frac{\Lambda}{\pi^2} =  \frac{M}{8}
\Leftrightarrow \ffbox{\frac{1}{g_0} = \frac{\Lambda}{\pi^2} + \frac{M}{8}}
\]
so that the two-point vertex obtains a non-trivial limit
\[
N g_0 \frac{M + \sqrt{p^2}}{8}
\]
Then, the propagator of $\alpha$ becomes
\[
\frac{8}{N g_0} \frac{1}{M + \sqrt{p^2}}
\]
and the four-point interaction of the fermions is given by
(Fig.~\ref{GN-interaction}) 
\[
- g_0 \frac{8}{N g_0} \frac{1}{M + \sqrt{p^2}}
= - \frac{8}{N} \frac{1}{M + \sqrt{p^2}} \stackrel{M\to
  0}{\longrightarrow} - \frac{8}{N} \frac{1}{\sqrt{p^2}}
\]

Hence, we have found
\[
\frac{1}{g_{0,cr}} = \frac{\Lambda}{\pi^2},\quad
y_E = 1,\quad x_h = 1
\]
and $g_E = \frac{M}{8} > 0$.

\subsection{Broken phase $g_0 > g_{0,cr}$}

In the broken phase, the maximum of $f[\alpha]$ takes place at a
non-vanishing point:
\[
\alpha = \frac{m}{\sqrt{g_0}}
\]
where $|m|$ is the mass of the fermions.  Shifting and rescaling $\alpha$
\[
\alpha \to \frac{m}{\sqrt{g_0}} + \frac{\alpha}{\sqrt{N}}
\]
we obtain
\[
S = - \int d^3 x \, \left[ \bar{\psi}^I \frac{1}{i} \sigma
  \cdot \partial \psi^I + i m \bar{\psi}^I \psi^I + \frac{1}{2}
  \alpha^2 + i \sqrt{\frac{g_0}{N}} \alpha\, \bar{\psi}^I \psi^I 
+ i \frac{N m}{g_0} \alpha \right]
\]
Now, the condition that $\alpha$ has zero VEV gives
\begin{eqnarray*}
i N \frac{m}{g_0} &=& \vev{\bar{\psi}^I \psi^I}\\
&=& - N \int_p\Tr \frac{1}{\sigma \cdot p + i m}\\
&=& i m \frac{N}{\pi^2} \left( \Lambda - \frac{\pi}{2} \sqrt{m^2} \right)
\end{eqnarray*}
Choosing $m > 0$, we must find
\[
\frac{1}{g_0} = \frac{\Lambda}{\pi^2} - \frac{m}{2 \pi} < \frac{1}{g_{0,cr}}
\]
This gives the same critical $g_{0,cr}$ and exponent $y_E=1$.  In this
phase $g_E = - \frac{m}{2 \pi} < 0$.

The two-point interaction vertex of $\alpha$ is given by
\begin{eqnarray*}
&&g_0 \int_q \Tr \frac{1}{\sigma \cdot (p+q) + im} \frac{1}{\sigma \cdot
  q + im}\\
&& = \frac{g_0}{\pi^2} \left( \Lambda - \frac{\pi}{2} m  - \pi \sqrt{p^2}
  \left( \frac{m^2}{p^2} + \frac{1}{4} \right) \arctan
  \frac{\sqrt{p^2}}{2m} \right)\\
&& = 1 - \frac{g_0}{\pi} \sqrt{p^2} \left( \frac{m^2}{p^2} +
  \frac{1}{4} \right) \arctan \frac{\sqrt{p^2}}{2m} 
\end{eqnarray*}
Hence, to leading order in $\frac{1}{N}$ the propagator of $\alpha$
becomes
\[
\frac{1}{\frac{g_0}{\pi} \sqrt{p^2} \left( \frac{m^2}{p^2} +
  \frac{1}{4} \right) \arctan \frac{\sqrt{p^2}}{2m}}
\]
and the 4-point interaction of the fermions is given by
\begin{eqnarray*}
- \frac{g_0}{N} \frac{1}{ \frac{g_0}{\pi} \sqrt{p^2} \left(
    \frac{m^2}{p^2} + \frac{1}{4} \right) \arctan
  \frac{\sqrt{p^2}}{2m} } &=& - \frac{\pi}{N} \frac{\sqrt{p^2}}{
\left( m^2 + \frac{p^2}{4} \right) \arctan
\sqrt{\frac{p^2}{4 m^2}} } \\
&\stackrel{m \to 0}{\longrightarrow}&
- \frac{8 \pi}{N} \frac{1}{\sqrt{p^2}}
\end{eqnarray*}

\newpage
\appendix

\section{Alternative: $\phi^4$ on a cubic lattice\label{alternative}}

The continuum limit of the Ising model in D=3, discussed at the end of
Lecture 1, can can be obtained from a different model. Let
$\phi_{\vec{n}}$ be a real variable taking a value from $- \infty$ to
$\infty$.  We consider the $\phi^4$ theory on a cubic lattice:
\[
S = \sum_{\vec{n}} \left[
\frac{1}{2} \sum_{i=1}^3 \left( \phi_{\vec{n}+\hat{i}} -
\phi_{\vec{n}} \right)^2 + \frac{m_0^2}{2} \phi_{\vec{n}}^2 +
\frac{\lambda_0}{4!} \phi_{\vec{n}}^4 \right]
\]
where $\lambda_0 > 0$, but $m_0^2$ can be negative.

For $\lambda_0$ fixed, the theory has two phases depending on the
value of $m_0^2$:
\begin{itemize}
\item \textbf{symmetric phase} $m_0^2 > m_{0, cr}^2 (\lambda_0)$:
  $\mathbf{Z}_2$ is intact.
\item \textbf{broken phase} $m_0^2 < m_{0, cr}^2 (\lambda_0)$:
  $\mathbf{Z}_2$ is spontaneously broken, and $\vev{\phi_{\vec{n}}}
  \ne 0$.
\end{itemize}
Note that the critical value $m_{0, cr}^2 (\lambda_0)$ depends on
$\lambda_0$.

The continuum limit is obtained as
\[
\vev{\phi (\vec{r}) \phi (\vec{0})}_{g; \mu} \equiv \mu \lim_{t \to
\infty} \e^{ (1+\eta) t} \vev{\phi_{\vec{n} = \mu \vec{r} \e^t}
\phi_{\vec{0}}}_{m_0^2 = m_{0, cr}^2 (\lambda_0) + \e^{- y_E t}
\frac{g}{\mu^2}}
\]
where $y_E, \eta$ are the same critical exponents as in the Ising
model.  This limit is not necessarily independent of $\lambda_0$.  For
independence, we need to rescale both $\phi$ and $g$:
\[
\left\lbrace\begin{array}{c@{~\longrightarrow~}c}
 \phi & \sqrt{z (\lambda_0)} ~\phi\\
 g & z_m (\lambda_0)~g
\end{array}
\right.
\]
and define
\[
\vev{\phi (\vec{r}) \phi (\vec{0})}_{g; \mu} \equiv z (\lambda_0) \mu
\lim_{t \to \infty} \e^{ (1+\eta) t} \vev{\phi_{\vec{n} = \mu \vec{r}
\e^t} \phi_{\vec{0}}}_{m_0^2 = m_{0, cr}^2 (\lambda_0) + \e^{- y_E t}
z_m (\lambda_0) \frac{g}{\mu^2}}
\]
\textbf{Universality} consists of two statements:
\begin{enumerate}
\item $y_E, \eta$ are the same as in the Ising model.
\item The continuum limit is the same as in the Ising model.  (We only
  have to choose $z(\lambda_0)$ and $z_m (\lambda_0)$ appropriately.)
\end{enumerate}

We have defined the continuum limit using lattice units for the
lattice theory.  How do we take the continuum limit if we use physical
units instead?  To use physical units, we assign the length
\[
a = \frac{1}{\Lambda} = \frac{1}{\mu e^t}
\]
to a lattice unit.  The action is now given by
\[
S = \underbrace{a^3 \sum_{\vec{n}}}_{= \int d^3 r} \Bigg[ \frac{1}{2}
  \sum_{i=1}^3 \underbrace{\frac{1}{a^2} \left( \varphi_{\vec{r} + a
  \hat{i}} - \varphi_{\vec{r}} \right)^2}_{= (\partial_i \varphi)^2} +
  \frac{m_{bare}^2}{2} \varphi_{\vec{r}}^2 + \frac{\lambda_{bare}}{4!}
  \varphi_{\vec{r}}^4 \Bigg]
\]
where
\[
\left\lbrace\begin{array}{c@{~\equiv~}l}
 \vec{r} & \vec{n} a = \frac{\vec{n}}{\Lambda}\\
 \varphi_{\vec{r}} & \frac{1}{\sqrt{a}} \phi_{\vec{n}} =
 \sqrt{\Lambda} \phi_{\vec{n}}\\
 m_{bare}^2 & \frac{m_0^2}{a^2} = m_0^2 \Lambda^2\\
 \lambda_{bare} & \frac{\lambda_0}{a} = \lambda_0 \Lambda
\end{array}\right.
\]
Then, to obtain the continuum limit we must choose
\[
\left\lbrace\begin{array}{c@{~=~}l} m_{bare}^2 & \Lambda^2 m_{0,cr}^2
 (\lambda_0) + z_m (\lambda_0) g \left( \frac{\Lambda}{\mu} \right)^{2
 - y_E}\\ \lambda_{bare} & \Lambda \lambda_0
\end{array}\right.
\]
and we obtain
\[
\vev{\phi (\vec{r}) \phi (\vec{0})}_{g; \mu}
 = z (\lambda_0) \lim_{\Lambda \to \infty}
 \left( \frac{\Lambda}{\mu} \right)^\eta \vev{ \varphi_{\vec{r}}
 \varphi_{\vec{0}} }_{m_{bare}^2, \lambda_{bare}}
\]
Note that $\lambda_0 > 0$ is an arbitrary finite constant.  The bare
squared mass has not only a quadratic divergence but also a divergence
of power $2 - y_E \simeq 0.4$.  The bare coupling is linearly
divergent.  We see clearly that \textbf{the UV divergences of
  parameters are due to the use of physical units.}  

\newpage
\section{Equivalence with linear models}

\subsection{Linear and non-linear $\sigma$ models}

The O(N) linear $\sigma$ model is given by
\[
S = - \int d^3 x\, \left[ \frac{1}{2} \partial_\mu \phi^I \partial_\mu
  \phi^I + \frac{m_0^2}{2} \phi^I \phi^I + \frac{\lambda}{8 N}
  \left( \phi^I \phi^I \right)^2 \right]
\]
where the self-coupling $\lambda$ has mass dimension 1, and the
model is perturbatively renormalizable.  We wish to show that the
continuum limit of the O(N) non-linear $\sigma$ model can be also
obtained as a limit of the renormalized linear $\sigma$ model.

As discussed in Appendix A, the continuum limit of the non-linear
$\sigma$ model can be constructed using the linear $\sigma$ model.
For given $\lambda_0$, there is a critical point $m^2_{0,cr}$ which
separates the symmetric and broken phases.  We can tune $m^2_0$ to
$m^2_{0,cr}$ to construct the same continuum limit.  But this is not
what we would like to do here.  

The linear $\sigma$ model is perturbatively renormalizable.  The
renormalized theory has two parameters: $\lambda$ and the
renormalized squared mass $m^2$.  We wish to show how to tune $m^2$ \&
$\lambda$ to construct the continuum limit of the non-linear
$\sigma$ model.

\begin{figure}[t]
\begin{center}
\epsfig{file=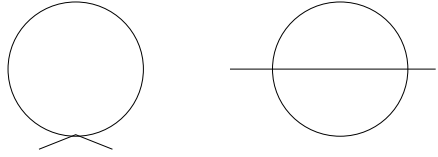, height=2cm}
\caption{UV divergent graphs\label{phi4}}
\end{center}
\end{figure}

The linear $\sigma$ model is superrenormalizable, and the only UV
divergent graphs are for the 1- \& 2-loop corrections to the squared
mass of the scalar (Fig.~\ref{phi4}).
For renormalization, we must take
\[
m_0^2 = A_N \lambda \Lambda - C_N \lambda^2 \ln \Lambda/\mu + m^2
\]
where $\mu$ is an arbitrary renormalization scale, and the constants
$A_N, C_N$ are given by
\[
A_N = - \frac{1}{4 \pi^2} \left( 1 + \frac{1}{2N} \right),\quad
C_N = - \frac{1}{32 \pi^2} \frac{1}{N} \left( 1 + \frac{2}{N} \right)
\]
Imposing that $m_0^2$ be invariant under a change of $\mu$, we obtain
the RG equation
\[
- \mu \frac{\partial m^2}{\partial \mu} = C_N \lambda^2
\]
This implies the following:
\begin{enumerate}
\item The dimensionless constant
\[
R_N \equiv \frac{m^2}{\lambda^2} - C_N \ln \frac{\lambda}{\mu}
\]
is an RG invariant.
\item The correlation functions satisfy the RG equation
\[
\vev{\phi^I (r) \phi^J (0)}_{m^2 + C_N \lambda^2 t,\, \lambda;\,\mu
  \e^{-t}} = \vev{\phi^I (r) \phi^J (0)}_{m^2,\, \lambda;\, \mu}
\]
\end{enumerate}
Since $m^2, \lambda, \phi$ have the mass dimensions $2, 1,
\frac{1}{2}$, respectively, the dimensional analysis gives
\[
\vev{\phi^I (r \e^{-t}) \phi^J (0)}_{m^2 \e^{2 t},\, \lambda \e^t;\,
  \mu \e^t} = \e^t \vev{\phi^I (r) \phi^J (0)}_{m^2,\, \lambda;\, \mu}
\]
Hence, the above RG equation can be rewritten as
\[
\vev{\phi^I (r \e^{-t}) \phi^J (0)}_{ \e^{2t} \left( m^2 + C_N
    \lambda^2 t \right),\, \lambda \e^t;\, \mu}
= \e^t \vev{\phi^I (r) \phi^J (0)}_{m^2,\, \lambda;\, \mu}
\]

Let us now consider the phase structure of the renormalized theory
parameterized by $m^2$, $\lambda$.  For given $\lambda$, we expect 
\begin{enumerate}
\item $m^2 < m_{cr}^2 (\lambda)$ in the broken phase
\item $m^2 > m_{cr}^2 (\lambda)$ in the symmetric phase
\end{enumerate}
The critical value $m_{cr}^2 (\lambda)$ depends on $\lambda$ (and
$\mu$).  Whether the system is in the broken phase or in the symmetric
phase should be RG invariant.  Hence, there should be a critical value
$R_{N,cr}$ of the RG invariant $R_N$ such that
\begin{enumerate}
\item $R_N < R_{N,cr}$ in the broken phase 
\item $R_N > R_{N,cr}$ in the symmetric phase 
\end{enumerate}
$R_{N, cr}$ is just a number that depends on $N$.  Hence, given
$\lambda$, the critical value of $m^2$ is given by
\[
m^2_{cr} (\lambda) \equiv \lambda^2 \left( R_{N,cr} + C_N \ln
  \frac{\lambda}{\mu} \right)
\]

Thus, making an analogy to the results of Appendix A, we expect that
the continuum limit of the non-linear $\sigma$ model is obtained as
\[
\vev{\phi^I (\vec{r}) \phi^J (0)}_{g_E; \mu}
= z_\lambda \lim_{t \to \infty} \e^{(1+\eta)t} \vev{\phi^I (\vec{r} \e^t) \phi^J
  (0)}_{m^2 = m^2_{cr} (\lambda) + g_E \e^{- y_E t},\, \lambda; \,\mu}
\]
where we choose $z_\lambda$ such that the limit does not depend on the
choice of $\lambda$. Using the RG equation, we can also write the
above as
\begin{eqnarray*}
\vev{\phi^I (\vec{r}) \phi^J (0)}_{g_E; \mu}
&=& z_\lambda \lim_{t \to \infty} \e^{\eta t} \vev{\phi^I (\vec{r}) \phi^J
  (0)}_{\e^{2t} (m^2_{cr} (\lambda)+C_N \lambda^2 t) + g_E
  \e^{(2-y_E)t},\, \lambda \e^t;\, \mu}\\
&=& \frac{z_\lambda}{\lambda^\eta} \lim_{t \to \infty} (\lambda
\e^t)^\eta \vev{\phi^I (\vec{r}) 
  \phi^J (0)}_{m^2_{cr} (\lambda \e^t) + \frac{g_E}{\lambda^{2-y_E}} (\lambda
  \e^t)^{2-y_E},\, \lambda \e^t;\, \mu}
\end{eqnarray*}
Hence, choosing 
\[
z_\lambda = \left(\frac{\lambda}{\mu}\right)^\eta
\]
and rescaling $g_E$ so that
\[
\frac{g_E}{\lambda^{2-y_E}} \longrightarrow \frac{g_E}{\mu^{2-y_E}}
\]
we obtain
\[
\vev{\phi^I (\vec{r}) \phi^J (0)}_{g_E; \mu}
= \frac{1}{\mu^\eta} \lim_{\lambda \to \infty} \lambda^\eta \vev{\phi^I
(r) \phi^J (0)}_{m^2 = m_{cr}^2 (\lambda) + g_E
  \left(\frac{\lambda}{\mu}\right)^{2-y_E},\, \lambda;\,\mu}
\]
The above formula implies that the non-linear $\sigma$ model is
obtained from the renormalized linear $\sigma$ model in the limit of
a strong coupling and a large squared mass.

Let us verify the above expectations using the large N approximations.
We rewrite the action by using an auxiliary field:
\begin{eqnarray*}
S &=& - \int d^3 x\, \left[ \frac{1}{2} (\partial_\mu \phi^I)^2 +
  \frac{1}{2} m_0^2 \phi^I \phi^I + \frac{\lambda}{8 N} (\phi^I
  \phi^I)^2 + \frac{N}{2 \lambda} \left( \alpha + i
    \frac{\lambda}{2 N} \phi^I \phi^I \right)^2 \right]\\
&=& - \int d^3 x\, \left[ \frac{1}{2} (\partial_\mu \phi^I)^2 +
  \frac{1}{2} (m_0^2 + i 
\alpha) \phi^I \phi^I + \frac{N}{2 \lambda} \alpha^2 \right]
\end{eqnarray*}
Denoting the saddle point of the $\alpha$ integration as $- i s$, we
expand
\[
\alpha = - i s + \sqrt{\frac{\lambda}{N}}\, \delta \alpha
\]
Substituting this into $S$, we obtain
\[
S = - \int d^3 x\, \left[ \frac{1}{2} (\partial_\mu \phi^I)^2 +
  \frac{\tilde{m}^2}{2} \phi^I \phi^I - i \sqrt{\frac{N}{\lambda}}\, s
  \delta \alpha + \frac{1}{2} (\delta \alpha)^2 +
  \sqrt{\frac{\lambda}{N}} \, i \delta \alpha \frac{1}{2} \phi^I
  \phi^I \right]
\]
where
\[
\tilde{m}^2 \equiv m_0^2 + s
\]
We only consider the O(N) symmetric phase, and take the physical
squared mass $\tilde{m}^2$ to be positive.

We can determine $s$ by the condition $\vev{\delta \alpha} = 0$.  To
leading order in 1/N, we obtain
\[
i \sqrt{\frac{N}{\lambda}} \, s - i \sqrt{\frac{\lambda}{N}}
\frac{N}{2} \int_p \frac{1}{p^2 + \tilde{m}^2} = 0
\]
Hence, we obtain
\[
s = \frac{\lambda}{2} \int_p \frac{1}{p^2 + \tilde{m}^2} =
\frac{\lambda}{4 \pi^2} \left( \Lambda - \tilde{m} \frac{\pi}{2}
\right)
\]
Therefore, we find
\[
m_0^2 = - s + \tilde{m}^2 = - \frac{\lambda}{4 \pi^2} \Lambda +
\underbrace{\tilde{m}^2 + \frac{\lambda}{8 \pi} \tilde{m}}_{= m^2}
\]
as expected from the values of $A_N, C_N$ quoted earlier.  At
criticality $\tilde{m}$ vanishes, and we find that the critical value
of $m^2$ vanishes to leading order in 1/N
\[
m_{cr}^2 (\lambda) =  0 \Longrightarrow R_{N,cr} = 0
\]
For large $\lambda$ and fixed $\tilde{m}$, we have
\[
m^2 \simeq \lambda \frac{\tilde{m}}{8 \pi}
\]
Hence,
\[
y_E = 1,\quad g_E = \frac{1}{8 \pi}\,\mu \tilde{m}
\]

In order to see the relation to the non-linear model, let us look at the
interaction.  The two-point interaction vertex of $\delta \alpha$
is given by
\[
\left( - i \sqrt{\frac{\lambda}{N}}\right)^2 \frac{N}{2} \int_q
\frac{1}{(q^2 + \tilde{m}^2)\left( (q+p)^2 + \tilde{m}^2 \right)} = -
\frac{\lambda}{8 \pi}  \frac{\arctan \sqrt{\frac{p^2}{4
      \tilde{m}^2}}}{\sqrt{p^2}} 
\]
Hence, the propagator of $\delta \alpha$ is given by
\[
\frac{1}{1 + \frac{\lambda}{8 \pi}  \frac{\arctan \sqrt{\frac{p^2}{4
      \tilde{m}^2}}}{\sqrt{p^2}}} 
\]
Thus, the four-point interaction vertex of the scalar field is given
by
\begin{eqnarray*}
\left( - i \sqrt{\frac{\lambda}{N}} \right)^2 \frac{1}{1 +
  \frac{\lambda}{8 \pi} \frac{\arctan \sqrt{\frac{p^2}{4
        \tilde{m}^2}}}{\sqrt{p^2}}}
&=& - \frac{8\pi}{N} \frac{\sqrt{p^2}}{\frac{8 \pi \sqrt{p^2}}{\lambda}
  + \arctan \sqrt{\frac{p^2}{4 \tilde{m}^2}}}\\
&\stackrel{\lambda \to \infty}{\longrightarrow}&
- \frac{8 \pi}{N} \frac{\sqrt{p^2}}{\arctan \sqrt{\frac{p^2}{4
      \tilde{m}^2}}}
\end{eqnarray*}
reproducing the large N limit obtained in sect.~2.

\newpage
\subsection{Yukawa and Gross-Neveu models}

The Yukawa model is given by
\[
S = - \int d^3 x \, \left[ \bar{\psi}^I \frac{1}{i} \sigma
  \cdot \partial \psi^I + \frac{1}{2} \partial_\mu \phi \partial_\mu
  \phi + \frac{M_0^2}{2} \phi^2 + i \frac{y}{\sqrt{N}} \phi
  \bar{\psi}^I \psi^I \right] 
\]
where the Yukawa coupling $y$ has dimension $\frac{1}{2}$, and the
model is perturbatively renormalizable.  We wish to show that the
continuum limit of the Gross-Neveu model can be also obtained as a
limit of the renormalized Yukawa model.

Let us first note the $\mathbf{Z_2}$ symmetry of the model.  The
action is invariant under
\[
\psi^I (x) \to \psi^I (-x),\quad
\bar{\psi}^I (x) \to - \bar{\psi}^I (-x),\quad
\phi (x) \to - \phi (-x)
\]
For given $y$, there is a critical value $M^2_{0,cr} (y^2)$ so
that the $\mathbf{Z_2}$ symmetry is broken spontaneously for $M_0^2 <
M^2_{0,cr} (y^2)$, and exact for $M_0^2 > M^2_{0,cr} (y^2)$.  By
universality, we can tune $M_0^2$ to $M^2_{0,cr}$ to obtain the same
continuum limit as that of the Gross-Neveu model.  We follow an
alternative route, and first renormalize the Yukawa model, and then
take a limit to obtain the continuum limit of the Gross-Neveu model.

\begin{figure}[t]
\begin{center}
\epsfig{file=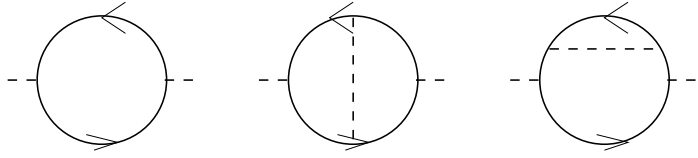, height=2cm}
\caption{UV divergences in the Yukawa model\label{yukawa}}
\end{center}
\end{figure}

The Yukawa model is also superrenormalizable, and the only UV
divergent graphs are for the 1- \& 2-loop corrections to the squared
mass of the scalar. (Fig.~\ref{yukawa})
We obtain
\[
M_0^2 = A_N y^2 \Lambda - C_N y^4 \ln \frac{\Lambda}{\mu} + M^2
\]
where
\[
A_N = \frac{1}{\pi^2},\quad C_N = \frac{3}{16 \pi^2} \frac{1}{N}
\]
The renormalized parameter $M^2$ satisfies
\[
- \mu \frac{\partial}{\partial \mu} M^2 = C_N y^4
\]
This implies the following:
\begin{enumerate}
\item The dimensionless constant
\[
R_N \equiv \frac{M^2}{y^4} - C_N \ln \frac{y^2}{\mu}
\]
is an RG invariant.
\item The correlation functions satisfy the RG equation
\[
\vev{\psi^I (r) \bar{\psi}^J (0)}_{M^2 + C_N y^4 t,\, y^2;\, \mu
  \e^{-t}}
= \vev{\psi^I (r) \bar{\psi}^J (0)}_{M^2,\, y^2;\, \mu}
\]
\end{enumerate}
Since $M^2, y^2, \psi^I$ have the scale dimensions $2, 1, 1$, we obtain
\[
\vev{\psi^I (r \e^{-t}) \bar{\psi}^J (0)}_{M^2 \e^{2t},\, y^2 \e^t;\,
  \mu \e^t} = \e^{2t} \vev{\psi^I (r) \bar{\psi}^J (0)}_{M^2,\, y^2;\,
  \mu}
\]
Using this, we can rewrite the RG equation as
\[
\vev{\psi^I (r \e^{-t}) \bar{\psi}^J (0)}_{\e^{2t} (M^2 + C_N y^4 t),\,
  \e^t y^2;\, \mu} = \e^{2 t} \vev{\psi^I (r) \bar{\psi}^J (0)}_{M^2,\, y^2;\,
  \mu}
\]

Let us now consider the phase structure of the renormalized theory.
For given $y^2$, we expect
\begin{enumerate}
\item $M^2 < M_{cr}^2 (y^2)$ or $R_N < R_{N,cr}$ in the broken phase
\item $M^2 > M_{cr}^2 (y^2)$ or $R_N > R_{N,cr}$ in the symmetric phase
\end{enumerate}
Here, $M_{0,cr}^2 (y^2)$ is given in terms of $R_{N,cr}$ as
\[
M_{cr}^2 (y^2) = y^4 \left( R_{N,cr} + C_N \ln \frac{y^2}{\mu} \right)
\]

Thus, we expect
\[
\vev{\psi^I (r) \bar{\psi}^J (0)}_{g_E;\, \mu}
= z_{y^2} \lim_{t \to \infty} \e^{(2+\eta)t} \vev{\psi^I (r \e^t)
  \bar{\psi}^J (0)}_{M^2 = M_{cr}^2 (y^2) + g_E \mu \e^{- y_E t},\,
  y^2;\, \mu}
\]
Using the RG equation, we can rewrite this as
\begin{eqnarray*}
\vev{\psi^I (r) \bar{\psi}^J (0)}_{g_E;\,\mu}
&=& z_{y^2} \lim_{t \to \infty} \e^{\eta t}\vev{\psi^I (r) \bar{\psi}^J
  (0)}_{\e^{2t} (M^2_{cr} (y^2) + C_N y^4 t) + g_E \mu \e^{(2-y_E)t},\,
  y^2 \e^t;\, \mu}\\
&=& \frac{z_{y^2}}{y^{2 \eta}} \lim_{t \to \infty} (y^2 \e^t)^{\eta} 
\vev{\psi^I (r) \bar{\psi}^J (0)}_{M^2_{cr} (y^2 \e^t) +
  \frac{g_E \mu}{y^{2(2-y_E)}} (y^2 \e^t)^{2-y_E} ,\, y^2 \e^t;\, \mu}
\end{eqnarray*}
Choosing
\[
z_{y^2} = \frac{y^{2 \eta}}{\mu^\eta}
\]
and redefining $g_E$ by
\[
\frac{g_E}{y^{2(2-y_E)}} \longrightarrow \frac{g_E}{\mu^{2-y_E}}
\]
we obtain
\[
\vev{\psi^I (r) \bar{\psi}^J (0)}_{g_E;\,\mu}
= \frac{1}{\mu^\eta} \lim_{y^2 \to \infty} 
y^{2 \eta} 
\vev{\psi^I (r) \bar{\psi}^J (0)}_{M^2_{cr} (y^2) +
  \frac{g_E}{\mu^{1-y_E}} y^{2(2-y_E)} ,\, y^2;\, \mu}
\]
This gives the continuum limit of the Gross-Neveu model in the limit
of large $M^2, y^2$.

Let us verify our expectations in the large N limit.  We consider only
the symmetric phase.  To the leading
order in 1/N, the scalar two-point interaction vertex is given by
\[
\left( \frac{- i y}{\sqrt{N}} \right)^2 N (-) \int_q \Tr
\frac{1}{\sigma \cdot q} \frac{1}{\sigma \cdot (q+p)}
= y^2 \left( \frac{\Lambda}{\pi^2} - \frac{\sqrt{p^2}}{8} \right)
\]
Hence, the scalar propagator is
\[
\frac{1}{p^2 + M_0^2 - y^2 \left( \frac{\Lambda}{\pi^2} -
    \frac{\sqrt{p^2}}{8} \right)}
\]
By tuning $M_0^2$ by
\[
M_0^2 = y^2 \frac{\Lambda}{\pi^2} + M^2
\]
the propagator becomes
\[
\frac{1}{p^2 + M^2 + y^2 \frac{\sqrt{p^2}}{8}}
\]
Thus, we obtain
\[
A_N = \frac{1}{\pi^2},\quad M_{cr}^2 = 0
\]
Hence, the interaction vertex of the fermions is given by
\[
- \frac{y^2}{N} \frac{1}{p^2 + M^2 + y^2 \frac{\sqrt{p^2}}{8}}
\]
For this to have a non-trivial limit as $y^2 \to \infty$, we must tune
\[
M^2 = y^2 g_E
\]
which implies $y_E = 1$.  Then, in the limit $y^2 \to \infty$ the
4-fermion vertex becomes
\[
- \frac{1}{N} \frac{1}{g_E + \frac{\sqrt{p^2}}{8}}
\]
reproducing the result for the Gross-Neveu model.


\begin{thebibliography}{99}
\bibitem{Wilson:1973jj}
  K.~G.~Wilson and J.~B.~Kogut,
  %``The Renormalization group and the epsilon expansion,''
  Phys.\ Rept.\  {\bf 12}, 75 (1974).
  %%CITATION = PRPLC,12,75;%%
\bibitem{Sonoda:2006rr}
  H.~Sonoda,
  %``Wilson's renormalization group and its applications in perturbation
  %theory,''
  arXiv:hep-th/0603151.
  %%CITATION = HEP-TH/0603151;%%
\bibitem{Coleman:1985as}
  S.~Coleman,
  \textit{Aspects of Symmetry}
  (Cambridge Univ. Press, 1985), Chapter~8
\bibitem{Arefeva:1979bd}
  I.~Y.~Arefeva,
  %``Phase Transition In The Three-Dimensional Chiral Field,''
  Annals Phys.\  {\bf 117}, 393 (1979).
\bibitem{Arefeva:1978fj}
  I.~Y.~Arefeva, E.~R.~Nissimov, and S.~J.~Pacheva,
  %``Bphzl Renormalization Of 1/N Expansion And Critical Behavior Of The
  %Three-Dimensional Chiral Field,''
  Commun.\ Math.\ Phys.\  {\bf 71}, 213 (1980).
\bibitem{Rosenstein:1988pt}
  B.~Rosenstein, B.~J.~Warr, and S.~H.~Park,
  %``The Four Fermi Theory Is Renormalizable in (2+1)-Dimensions,''
  Phys.\ Rev.\ Lett.\  {\bf 62}, 1433 (1989).
  %%CITATION = PRLTA,62,1433;%%
\end{thebibliography}
\end{document}